\documentstyle[aps,amssymb]{revtex}

\begin{document}
\author{R.A. Serota\thanks{%
serota@physics.uc.edu} and B. Goodman\thanks{%
goodman@physic.uc.edu}}
\address{Department of Physics\\
University of Cincinnati\\
Cincinnati, OH\ 45221-0011}
\title{Classical absorption in small metal particles and thin films}
\date{06/01/98}
\maketitle

\begin{abstract}
We study the electric dipole absorption in small metal particles and thin
films in a longitudnal electric field. In diffusive approximation, we give
both the phenomenological and microscopic derivations with the account for
Thomas-Fermi screening and Drude relaxation.
\end{abstract}

\section{Introduction}

Classical absorption in a variable electromagnetic field is a well studied
and documented field of research\cite{LL}. Here, we will be primarily
concerned with the absorption of the longitudinal component of a
''quasi-static'' electric field $E_{0}e^{i\omega t}$. This problem arises,
for instance, in study of absorption of electromagnetic radiation by
particles whose size is smaller than the wave-length $\lambda $\cite{LL},%
\cite{BH} or for absorption in a film sandwiched between the capacitor
plates. We point out the role of the diffusion, in addition to ohmic
conduction, in the dynamics of the charge transport.

Absorption is determined by the imaginary (''dynamical'') part of the
electric moment ${\cal P}e^{i\omega t}$ \cite{LL} 
\begin{equation}
Q=\frac{1}{2}%
\mathop{\rm Re}%
\left\{ i\omega {\cal P}E_{0}\right\} =-\frac{1}{2}\omega E_{0}%
\mathop{\rm Im}%
\left\{ {\cal P}\right\}  \label{Q}
\end{equation}
For metals, the standard approach is to characterize the particles by the
Drude dielectric constant, 
\begin{equation}
\epsilon _{D}=1-\frac{4\pi i\sigma _{D}}{\omega }  \label{epsilon_Drude}
\end{equation}
where 
\begin{equation}
\sigma _{D}=\frac{\sigma _{0}}{1+i\omega \tau }  \label{sigma_Drude}
\end{equation}
is the Drude conductivity and $\sigma _{0}$ and $\tau $ are the Boltzmann
conductivity and the scattering time respectively.

At small frequencies, $\epsilon _{D}$ is dominated by the imaginary part.
Consequently, that the quasi-static polarization $P_{st}\thicksim E_{0}/4\pi 
$, attributed to the surface charge, screens the external field and produces
the electric moment 
\[
{\cal P}_{st}=P_{st}V\thicksim \frac{E_{0}V}{4\pi }
\]
where $V$ is the volume. The dynamical polarization $P_{\omega }\thicksim $ $%
-E_{0}/4\pi \epsilon _{D}$ produces the electric moment 
\[
{\cal P}_{\omega }\thicksim -\frac{i\omega E_{0}V}{\left( 4\pi \right)
^{2}\sigma _{0}}
\]
which, upon substitution in eq. (\ref{Q}), yields the standard
Rayleigh-Drude absorption 
\[
Q_{RD}\thicksim \frac{\left( \omega E_{0}\right) ^{2}V}{2\left( 4\pi \right)
^{2}\sigma _{0}}
\]

Another way to interpret this result is to notice that the current density
in the bulk is given by 
\[
j_{\omega }\thicksim i\omega P_{st}\thicksim \frac{i\omega E_{0}}{4\pi } 
\]
This current is responsible for the surface charge that screens the external
field. It couples to the dynamical field in the bulk 
\[
E_{\omega }\thicksim \frac{j_{\omega }}{\sigma _{0}} 
\]
resulting in absorption 
\[
Q=\frac{1}{2}%
\mathop{\rm Re}%
\left\{ j_{\omega }E_{\omega }^{*}\right\} V=Q_{RD} 
\]

The surface charge is, of course, only a mathematical construction. In
reality, the field is screened over the screening length $\Lambda ^{-1}$,
where $\Lambda $ is the Thomas-Fermi wave-vector 
\begin{equation}
\Lambda ^{2}=4\pi e^{2}\frac{dn}{d\mu }=\frac{4\pi \sigma _{0}}{D}
\label{Lambda}
\end{equation}
$dn/d\mu $ being the thermodynamic density of states and $D$ is the
diffusion constant. For a good metal $\Lambda ^{-1}$ is very small -
typically an Angstrom - and the sample size $L$ is such that $L\Lambda \gg 1$
. Furthermore, $\ell \Lambda \gg 1$ where $\ell =\sqrt{D\tau }$ is the
electron mean-free-path. In semiconductors, on the other hand, various
relationships between $L$, $\ell $ and $\Lambda $ are possible. The Drude
dielectric constant obviously cannot describe the charge distribution of the
screening layer. This paper presents a natural extension of the classical
argument based on $\sigma _{D}$ alone to incorporate both static and
dynamical aspects of the finite thickness of the screening layer.

\section{Phenomenological description of absorption}

\subsection{Maxwell equations and boundary conditions}

Consider a sample in the quasi-static field $E_{z}=E_{0}e^{i\omega t}$.
Using the Coulomb gauge for the longitudinal field, we solve together the
Maxwell equation and the continuity condition

\begin{eqnarray}
4\pi \rho \left( {\bf r};\omega \right) &=&{\bf \nabla \cdot E}\left( {\bf r}%
;\omega \right)  \label{Maxwell_eq} \\
-i\omega \rho \left( {\bf r};\omega \right) &=&{\bf \nabla \cdot j}\left( 
{\bf r};\omega \right)  \label{Continuity_condition}
\end{eqnarray}
Eliminating $\rho $, we find 
\[
{\bf \nabla \cdot }\left( {\bf j\left( r;\omega \right) +}\frac{i\omega }{%
4\pi }{\bf E}\left( {\bf r};\omega \right) \right) =0 
\]
whereof the current density can be written as 
\begin{equation}
{\bf j}\left( {\bf r};\omega \right) =\frac{i\omega }{4\pi }\left( {\bf D}%
\left( {\bf r};\omega \right) -{\bf E}\left( {\bf r};\omega \right) \right)
\label{j_solution}
\end{equation}
in terms of the divergenceless vector ${\bf D}$ 
\begin{equation}
{\bf \nabla \cdot D}\left( {\bf r};\omega \right) =0  \label{D_eq}
\end{equation}

The boundary conditions complementing eq. (\ref{j_solution}) can be obtained
as follows. First, there is no current through the sample boundary $\partial 
$ 
\begin{equation}
j_{n}\left( {\bf r};\omega \right) |_{\partial }=0  \label{j_boundary}
\end{equation}
and, consequently, 
\begin{equation}
D_{n}\left( {\bf r};\omega \right) |_{\partial }=E_{n}\left( {\bf r};\omega
\right) |_{\partial }  \label{D_boundary_condition}
\end{equation}
Second, since $\rho $ is bounded and there is no surface charge, the
electric field must be continuous at the boundary 
\begin{equation}
{\bf E}\left( {\bf r};\omega \right) |_{\partial }={\bf E}^{out}\left( {\bf r%
};\omega \right) |_{\partial }  \label{E_boundary_condition}
\end{equation}
where ${\bf E}^{out}$ is the field outside the sample.

${\bf D}$ has a meaning of electric displacement field for a ''dielectric''
medium in which all the current is assigned to the polarization ${\bf P}$,
namely, 
\begin{equation}
{\bf j}\left( {\bf r};\omega \right) =i\omega {\bf P}\left( {\bf r};\omega
\right)  \label{j_vs_P}
\end{equation}
so that 
\begin{equation}
{\bf D}\left( {\bf r};\omega \right) ={\bf E}\left( {\bf r};\omega \right)
+4\pi {\bf P}\left( {\bf r};\omega \right)  \label{D_full}
\end{equation}
and 
\begin{equation}
{\bf \nabla \cdot P}\left( {\bf r};\omega \right) =-\rho \left( {\bf r}%
;\omega \right)  \label{P_vs_rho}
\end{equation}
Eq. (\ref{j_solution}) can be used also to define ${\bf D}^{out}={\bf E}%
^{out}$ outside the body. From eqs. (\ref{E_boundary_condition}) and (\ref
{D_boundary_condition}) it then follows that the normal component of ${\bf D}
$ would be continuous at the boundary. On the other hand, the tangential
component of ${\bf D}$ may be discontinuous due to the tangential current at
the boundary.

\subsection{Calculation of the absorption}

Local absorption is given by 
\begin{equation}
Q\left( {\bf r};\omega \right) =\frac{1}{2}%
\mathop{\rm Re}%
\left\{ {\bf j}\left( {\bf r};\omega \right) {\bf \cdot E}^{*}\left( {\bf r}%
;\omega \right) \right\} =-\frac{1}{2}%
\mathop{\rm Im}%
\left\{ \omega {\bf P}\left( {\bf r};\omega \right) {\bf \cdot E}^{*}\left( 
{\bf r};\omega \right) \right\}   \label{Q_local-1}
\end{equation}
Using eq. (\ref{j_solution}), the latter can be also written as 
\begin{equation}
Q\left( {\bf r};\omega \right) =-%
\mathop{\rm Im}%
\left\{ \frac{\omega }{8\pi }{\bf D}\left( {\bf r};\omega \right) {\bf \cdot %
E}^{*}\left( {\bf r};\omega \right) \right\}   \label{Q_local-2}
\end{equation}
It must be emphasized that in evaluation of the total absorption $Q$ 
\begin{equation}
Q=\int Q\left( {\bf r};\omega \right) d{\bf r}  \label{Q_total}
\end{equation}
{\it \ the actual field in the interior of the sample can be replaced by the
applied field} (i.e. in the absence of the sample). For instance, using eqs.
(\ref{D_eq}) and (\ref{D_boundary_condition}), ${\bf E}^{*}\left( {\bf r}%
;\omega \right) {\bf =-\nabla }\phi ^{*}\left( {\bf r};\omega \right) $ and
the Gauss theorem, we find the total absorption as 
\begin{equation}
Q=\frac{\omega }{2}%
\mathop{\rm Im}%
\left\{ \int \rho \left( {\bf r};\omega \right) \phi ^{*}\left( {\bf r}
;\omega \right) d{\bf r}\right\}   \label{Q_total-1}
\end{equation}
It is shown in Appendix A that the latter can be also written 
\begin{equation}
Q=\frac{\omega }{2}%
\mathop{\rm Im}%
\left\{ \int \rho \left( {\bf r};\omega \right) \phi _{0}\left( {\bf r}%
\right) d{\bf r}\right\} =-\frac{\omega }{2}%
\mathop{\rm Im}%
\left\{ {\bf E}_{0}{\bf \cdot }\overrightarrow{{\cal P}}\right\} 
\label{Q_total-2}
\end{equation}
where 
\begin{equation}
\phi _{0}\left( {\bf r}\right) =-{\bf E}_{0}{\bf \cdot r}  \label{phi_0}
\end{equation}
is the potential of the applied field and 
\begin{equation}
\overrightarrow{{\cal P}}=\int \rho \left( {\bf r};\omega \right) {\bf r}d%
{\bf r}  \label{Pcal}
\end{equation}
is the electric moment of the system (of course, the second eq. (\ref
{Q_total-2}) also follows from a general thermodynamic argument \cite{LL}).
The physical reason why either $\phi $ or $\phi _{0}$ can be used in the
dissipation integral is that their difference is due to the screening charge
itself and is the (non-dissipative) energy of interaction of the screening
charge with itself.

By means of the Gauss's theorem and eqs. (\ref{P_vs_rho}), (\ref{j_boundary}%
) and (\ref{j_vs_P}), we also find 
\[
{\bf E}_{0}{\bf \cdot }\overrightarrow{{\cal P}}=\int -\left[ {\bf \nabla
\cdot }\left( {\bf P}\left( {\bf r};\omega \right) \left( {\bf E}_{0}{\bf 
\cdot r}\right) \right) -{\bf E}_{0}{\bf \cdot P}\left( {\bf r};\omega
\right) \right] d{\bf r=E}_{0}{\bf \cdot }\int {\bf P}\left( {\bf r};\omega
\right) d{\bf r} 
\]
that is 
\begin{equation}
\overrightarrow{{\cal P}}{\bf =}\int {\bf P}\left( {\bf r};\omega \right) d%
{\bf r}  \label{Pcal_vs_P}
\end{equation}
which is consistent with the fact that ${\bf E}$ can be replaced by ${\bf E}%
_{0}$ in the evaluation of $Q$.

\subsection{Constitutive equation and solution for the field}

A\ constitutive equation is needed to complete the description of the
medium. For this we use the\ ''generalized Einstein transport equation'' 
\begin{equation}
{\bf j}\left( {\bf r};\omega \right) =\sigma _{D}{\bf E}\left( {\bf r}%
;\omega \right) -D_{D}{\bf \nabla }\rho \left( {\bf r};\omega \right)
\label{Einstein_rel}
\end{equation}
where 
\begin{equation}
D_{D}=\frac{D}{1+i\omega \tau }  \label{D_Drude}
\end{equation}
Eq. (\ref{Einstein_rel}) will be further discussed in Section III. Combining
eqs. (\ref{j_solution}) and (\ref{Einstein_rel}), we find, with the help of
eq. (\ref{Maxwell_eq}) the following equation for the electric field 
\begin{equation}
{\bf \nabla }^{2}{\bf E}\left( {\bf r};\omega \right) -\widetilde{\Lambda }%
^{2}{\bf E}\left( {\bf r};\omega \right) =-\frac{i\omega }{D_{D}}{\bf D}%
\left( {\bf r};\omega \right)  \label{E_eq}
\end{equation}
Here $\widetilde{\Lambda }$ is the dynamical screening length given by 
\begin{equation}
\widetilde{\Lambda }^{2}=\frac{i\omega \epsilon _{D}}{D_{D}}=\Lambda
^{2}\left( 1+\frac{i\omega }{4\pi \sigma _{D}}\right)  \label{Lambda_complex}
\end{equation}
Solution of eq. (\ref{E_eq}) is found as 
\begin{equation}
{\bf E}\left( {\bf r};\omega \right) =\frac{{\bf D}\left( {\bf r};\omega
\right) }{\epsilon _{D}}+{\bf E}^{h}\left( {\bf r};\omega \right)
\label{E_solution}
\end{equation}
where ${\bf E}^{h}$ is the solution of the homogeneous equation 
\begin{equation}
{\bf \nabla }^{2}{\bf E}^{h}\left( {\bf r};\omega \right) -\widetilde{%
\Lambda }^{2}{\bf E}^{h}\left( {\bf r};\omega \right) =0  \label{E_hom_eq}
\end{equation}
such that the boundary condition (\ref{E_boundary_condition}) is satisfied.
Note that ${\bf E}^{h}$ represents the deviation from the standard
Rayleigh-Drude description.

The procedure for finding the electric field can be, in general, described
as follows. First, one determines the displacement vector ${\bf D=}-{\bf %
\nabla }\phi ^{d}$ , 
\begin{eqnarray}
\nabla ^{2}\phi ^{d} &=&0  \label{phi_D_equation} \\
\nabla _{n}\phi ^{d}|_{\partial } &=&-E_{n}^{out}|_{\partial }
\label{phi_D_bc}
\end{eqnarray}
as per eqs. (\ref{D_eq}) and (\ref{D_boundary_condition}), and the solution
of the homogeneous equation 
\begin{eqnarray}
\nabla ^{2}\phi ^{h}\left( {\bf r};\omega \right) -\widetilde{\Lambda }%
^{2}\phi ^{h}\left( {\bf r};\omega \right)  &=&0  \label{phi_h_equation} \\
\nabla _{n}\phi ^{h}|_{\partial } &=&-E_{n}^{out}|_{\partial }\left( \frac{%
\epsilon _{D}-1}{\epsilon _{D}}\right)   \label{phi_h_bc}
\end{eqnarray}
as per eqs. (\ref{E_hom_eq}), (\ref{E_solution}) and ${\bf E}^{h}=-{\bf %
\nabla }\phi ^{h}$. Subsequently, the total field in the interior is found
as 
\begin{eqnarray}
\phi  &=&\phi ^{h}+\frac{\phi ^{d}}{\epsilon _{D}}  \label{phi_equation} \\
\phi |_{\partial } &=&-\phi ^{out}|_{\partial }  \label{phi_bc}
\end{eqnarray}
as per eqs. (\ref{E_solution}) and (\ref{E_boundary_condition}).

Clearly, the use of eq. (\ref{Einstein_rel}) assumes the applicability of
the diffusion approximation. The conditions for the latter are $\omega
\lesssim \tau ^{-1}$ and, strictly speaking, $\ell \Lambda \ll 1$. For $\ell
\Lambda \gg 1$, eq. (\ref{Einstein_rel}) can still be used in the bulk. The
situation within the screening layer at the boundary must be analyzed
separately.

\subsection{Examples: fields and absorption in a slab and sphere}

\subsubsection{Solution for a slab}

Consider a narrow slab, $L_{z}\equiv L\ll L_{x},L_{y}\thicksim \sqrt{S}$.
Transverse degrees of freedom can be neglected so the only variable is $z$
and all fields are directed along $z$. The boundary condition is that the
electric field inside and outside the slab match at the boundary. This is
because the field of the slab's electric moment is zero, ${\cal E}=0$.
Indeed, this field has to satisfy the equation $\frac{d}{dz}{\cal E}=0$ (no
charge) and fall off to zero at infinity (it is also just a field of two
infinite plates of opposite charge). Thus $D=E^{out}=E_{0}$ and the boundary
condition is given by 
\begin{equation}
E(0;\omega )=E(L;\omega )=E_{0}  \label{bc-slab}
\end{equation}
Eqs. (\ref{j_solution}) and (\ref{E_eq}) become respectively 
\begin{equation}
j\left( z;\omega \right) =-\frac{i\omega }{4\pi }\left( E\left( z;\omega
\right) -E_{0}\right)   \label{j-slab}
\end{equation}
and: 
\begin{equation}
\frac{d^{2}}{dz^{2}}E\left( z;\omega \right) -\widetilde{\Lambda }%
^{2}E\left( z;\omega \right) =-\frac{i\omega }{D_{D}}E_{0}  \label{E_eq-slab}
\end{equation}
Solving eq. (\ref{E_eq-slab}), subject to the boundary condition (\ref
{bc-slab}), we find 
\begin{eqnarray}
E\left( z;\omega \right)  &=&\left( 1+\left( \epsilon _{D}-1\right) \frac{%
\cosh \left( \left( z-\frac{L}{2}\right) \widetilde{\Lambda }\right) }{\cosh 
\frac{L}{2}\widetilde{\Lambda }}\right) \frac{E_{0}}{\epsilon _{D}}
\label{E_solution-slab} \\
E_{st}(z) &=&E_{0}\frac{\cosh \left( \left( z-\frac{L}{2}\right) \Lambda
\right) }{\cosh \frac{L}{2}\Lambda }  \label{E_static_solution-slab}
\end{eqnarray}
and 
\begin{eqnarray}
\phi \left( z;\omega \right)  &=&\left( -\left( z-\frac{L}{2}\right) -\frac{
\left( \epsilon _{D}-1\right) }{\widetilde{\Lambda }}\frac{\sinh \left(
\left( z-\frac{L}{2}\right) \widetilde{\Lambda }\right) }{\cosh \frac{L}{2}%
\widetilde{\Lambda }}\right) \frac{E_{0}}{\epsilon _{D}}
\label{phi_solution-slab} \\
\phi _{st}(z) &=&-\frac{E_{0}}{\Lambda }\frac{\sinh \left( \left( z-\frac{L}{%
2}\right) \Lambda \right) }{\cosh \frac{L}{2}\Lambda }
\label{phi_static_solution-slab}
\end{eqnarray}
for the field and potential inside the slab, with the static components
explicitly emphasized.

Local absorption is given by 
\begin{equation}
Q\left( z\right) =\frac{1}{2}%
\mathop{\rm Re}%
\left\{ j\left( z;\omega \right) E^{*}\left( z;\omega \right) \right\} =%
\frac{1}{8\pi }\omega E_{0}%
\mathop{\rm Im}%
\left\{ E\left( z;\omega \right) \right\}   \label{Q_local-slab}
\end{equation}
The total absorption of a slab is 
\begin{eqnarray}
Q &=&S\int_{0}^{L}Q\left( z\right) dz=\frac{1}{8\pi }\omega
E_{0}S\int_{0}^{L}%
\mathop{\rm Im}%
\left\{ E\left( z;\omega \right) \right\} dz  \label{Q_total-slab} \\
&=&-\frac{1}{2}\omega E_{0}%
\mathop{\rm Im}%
\left\{ {\cal P}\right\}   \label{Q_vs_Pcal-slab}
\end{eqnarray}
where the latter expression - in agreement with eq. (\ref{Q_total-2}) - is
written in terms of the electric moment of the slab 
\begin{equation}
{\cal P}=S\int z\rho (z;\omega )dz=\frac{VE_{0}}{4\pi }-\frac{1}{4\pi }%
S\int_{0}^{L}E\left( z;\omega \right) dz  \label{Pcal-slab}
\end{equation}
and integration by parts has been used. Further integration of by parts
yields 
\begin{equation}
{\cal P}=\frac{VE_{0}}{4\pi }+\frac{1}{4\pi }S\left( \phi (L;\omega )-\phi
(0;\omega )\right) =\frac{VE_{0}}{4\pi }-\frac{1}{2\pi }S\phi (0;\omega )
\label{Pcal_vs_Phi-slab}
\end{equation}
and combining this with eq. (\ref{phi_solution-slab}), we obtain the
following expression for the electric moment of the slab: 
\begin{eqnarray}
{\cal P} &=&\frac{VE_{0}}{4\pi }\frac{\left( \epsilon _{D}-1\right) }{%
\epsilon _{D}}\left( 1-\frac{\tanh \left( L\widetilde{\Lambda }/2\right) }{L%
\widetilde{\Lambda }/2}\right)   \label{Pcal_solution-slab} \\
{\cal P}_{st} &=&\frac{VE_{0}}{4\pi }\left( 1-\frac{\tanh \left( L\Lambda
/2\right) }{L\Lambda /2}\right)   \label{Pcal_static_solution-slab}
\end{eqnarray}
where the static component is explicitly emphasized. Using eqs. (\ref
{Q_vs_Pcal-slab}) and (\ref{Pcal_solution-slab}), we find the closed form
solution for the total absorption in the slab 
\begin{equation}
Q=\frac{1}{8\pi }\omega E_{0}^{2}V%
\mathop{\rm Im}%
\left\{ \frac{\left( \epsilon _{D}-1\right) }{\epsilon _{D}}\left( 1-\frac{%
\tanh \left( L\widetilde{\Lambda }/2\right) }{L\widetilde{\Lambda }/2}%
\right) \right\}   \label{Q_solution-slab}
\end{equation}

Notice that {\it no assumption about the magnitude of }$L\Lambda ${\it \ has
been made}. For $L\Lambda \gg 1$, expanding eq. (\ref{Q_solution-slab}) at
small frequencies, $\omega \ll D/L^{2}$, we find 
\begin{equation}
Q=\frac{\left( \omega E_{0}\right) ^{2}V}{2\left( 4\pi \right) ^{2}\sigma
_{0}}\left( 1-\frac{3}{L\Lambda }\right) =Q_{RD}\left( 1-\frac{3}{L\Lambda }%
\right)  \label{Q_omega2_solution-slab}
\end{equation}
The expression in front of the parentheses is the standard Rayleigh-Drude
result $Q_{RD}$. The second term inside the parentheses is the screening
layer volume effect.

\subsubsection{Solution for a sphere}

The field outside the sphere of radius $a$, ${\bf E}^{out}=-{\bf \nabla }%
\phi ^{out}$, where\cite{LL} 
\begin{equation}
\phi ^{out}\left( {\bf r};\omega \right) =-{\bf E}_{0}{\bf \cdot r}+\frac{%
\overrightarrow{{\cal P}}{\bf \cdot r}}{r^{3}}  \label{phi_out-sphere}
\end{equation}
and $\overrightarrow{{\cal P}}\parallel {\bf E}_{0}$ is the sphere's
electric moment. $\phi ^{out}$ is proportional to the scalar ${\bf E}_{0}%
{\bf \cdot r\propto \cos }\left( \theta \right) $ and so is the interior
solution $\phi ^{d}$ which also obeys the Laplace equation. Matching
according to eq. (\ref{phi_D_bc}) gives 
\begin{equation}
\phi ^{d}\left( {\bf r};\omega \right) =-{\bf E}_{0}{\bf \cdot r}\left( 1+%
\frac{2{\cal P}}{E_{0}a^{3}}\right)   \label{phi_in-sphere}
\end{equation}
Consequently, the field inside the sphere is 
\begin{equation}
{\bf E}\left( {\bf r};\omega \right) =\frac{{\bf E}_{0}\left( 1+\frac{2{\cal %
P}}{E_{0}a^{3}}\right) }{\epsilon _{D}}+{\bf E}^{h}\left( {\bf r};\omega
\right)   \label{E_in-sphere}
\end{equation}
with the homogeneous solution found as ${\bf E}^{h}=-{\bf \nabla }\phi ^{h}$%
, where 
\begin{eqnarray}
\phi ^{h}\left( {\bf r};\omega \right)  &=&A\frac{a{\bf E}_{0}{\bf \cdot r}}{%
r}i_{1}\left( r\widetilde{\Lambda }\right)   \label{phi_hom-sphere} \\
i_{1}\left( x\right)  &=&\sqrt{\frac{\pi }{2x}}I_{\frac{3}{2}}\left(
x\right)   \label{BesselI_spherical_3/2}
\end{eqnarray}
and $i_{1}\left( x\right) $ and $I_{\frac{3}{2}}\left( x\right) $ is the
spherical Bessel and Bessel function of imaginary argument respectively. The
constants $A$ and ${\cal P}$ are found from the boundary conditions (\ref
{E_boundary_condition}) (or, equivalently, eqs.(\ref{phi_h_equation}) and (%
\ref{phi_equation})) and, using the notation 
\begin{equation}
X=1-3\frac{\coth \left( a\widetilde{\Lambda }\right) }{a\widetilde{\Lambda }}%
+\frac{3}{\left( a\widetilde{\Lambda }\right) ^{2}}  \label{X}
\end{equation}
we obtain the following expression for $A$ 
\[
A=\frac{3}{\epsilon _{D}+2X}\left( \epsilon _{D}-1\right) 
\]
and the electric moment of the sphere: 
\begin{eqnarray}
{\cal P} &=&\frac{3VE_{0}}{4\pi }\left( \frac{X}{\epsilon _{D}+2X}\left(
\epsilon _{D}-1\right) \right)   \label{phi_solution-sphere} \\
{\cal P}_{st} &=&\frac{3VE_{0}}{4\pi }\left( 1-3\frac{\coth \left( a\Lambda
\right) }{a\Lambda }+\frac{3}{\left( a\Lambda \right) ^{2}}\right) 
\label{Pcal_solution_static-sphere}
\end{eqnarray}
where the static component is explicitly emphasized. The latter coincides
with the result of Ref.\cite{LS} obtained quantum-mechanically.

Using eq. (\ref{Q}), we find the closed form solution for the absorption in
a sphere 
\begin{equation}
Q=\frac{3}{8\pi }\omega E_{0}^{2}V%
\mathop{\rm Im}%
\left\{ \left( \frac{X}{\epsilon _{D}+2X}\left( \epsilon _{D}-1\right)
\right) \right\}  \label{Q_solution-sphere}
\end{equation}
where, again, {\it no assumption about the magnitude of }$a\Lambda ${\it \
has been made}. For $a\Lambda \gg 1$, the small-frequency expansion,

\begin{equation}
Q=\frac{9\left( \omega E_{0}\right) ^{2}V}{2\left( 4\pi \right) ^{2}\sigma
_{0}}\left( 1-\frac{11}{2a\Lambda }\right) =Q_{RD}\left( 1-\frac{11}{%
2a\Lambda }\right)  \label{Q_omega2_solution-sphere}
\end{equation}
If the oscillating field is due to electromagnetic radiation with
wave-length longer than the sphere size, then dividing by the density of the
incoming flux find for effective cross-section of the absorption 
\begin{equation}
\Sigma =\frac{9\omega ^{2}V}{4\pi \sigma _{0}c}\left( 1-\frac{11}{2a\Lambda }%
\right)  \label{Sigma-sphere}
\end{equation}

\section{Response-function formalism}

Using standard Fermi-liquid considerations\cite{PN} in real space or a
field-theoretical argument\cite{SYK}, it can be shown that the linear
response of the current to the electric field can be written as 
\begin{eqnarray}
j_{\alpha }\left( {\bf r};\omega \right) &=&\int \sigma _{\alpha \beta
}\left( {\bf r},{\bf r}^{\prime };\omega \right) E_{\beta }\left( {\bf r}%
^{\prime };\omega \right) d{\bf r}^{\prime }  \label{j_linear_response} \\
\sigma _{\alpha \beta }\left( {\bf r},{\bf r}^{\prime };\omega \right)
&=&\sigma _{D}\left( \delta _{\alpha \beta }\delta ({\bf r}-{\bf r}^{\prime
})-{\bf \nabla }_{\alpha }{\bf \nabla }_{\beta }^{\prime }d\left( {\bf r},%
{\bf r}^{\prime };\omega \right) \right)  \label{sigma_tensor}
\end{eqnarray}
where $d$ is the diffusion propagator (''diffuson'')

\begin{equation}
{\bf \nabla }^{2}d\left( {\bf r},{\bf r}^{\prime };\omega \right) =-\delta (%
{\bf r}-{\bf r}^{\prime })+\frac{i\omega }{D_{D}}d\left( {\bf r},{\bf r}%
^{\prime };\omega \right)  \label{d_equation}
\end{equation}
satisfying the boundary condition

\begin{equation}
{\bf \nabla }_{n}d\left( {\bf r},{\bf r}^{\prime };\omega \right)
|_{\partial }={\bf \nabla }_{n}^{\prime }d\left( {\bf r},{\bf r}^{\prime
};\omega \right) |_{\partial ^{\prime }}=0  \label{d_boundary_condition}
\end{equation}

Clearly, the linear response described by eqs. (\ref{j_linear_response}) and
(\ref{sigma_tensor}) is the real-space variant of the non-local current
response described in Chapter III of Ref.\cite{PN}. In particular, for an
infinite homogeneous system in a longitudinal field $\widehat{{\bf q}}%
E_{0}\exp \left( i{\bf q}\cdot {\bf r}\right) $, it is in agreement with the
non-local ($q$-dependent) conductivity given by eq. (3.145)\ in Ref.\cite{PN}%
. It is also equivalent to the ''generalized Einstein equation'' (see eq. (%
\ref{Einstein_rel}) and below), the latter being its differential form.

Using eqs. (\ref{Maxwell_eq}), (\ref{Continuity_condition}), (\ref
{d_equation}) and (\ref{d_boundary_condition}) and ${\bf E=-\nabla }\phi $,
we find from eq. (\ref{j_linear_response}) via integration by parts,

\begin{equation}
{\bf j}\left( {\bf r};\omega \right) =-i\omega e^{2}\frac{dn}{d\mu }{\bf %
\nabla }\int d\left( {\bf r},{\bf r}^{\prime };\omega \right) \phi \left( 
{\bf r}^{\prime };\omega \right) d{\bf r}^{\prime }
\label{j_linear_response_2}
\end{equation}
and

\begin{eqnarray}
\rho \left( {\bf r};\omega \right) &=&\int \Pi \left( {\bf r},{\bf r}%
^{\prime };\omega \right) \phi \left( {\bf r}^{\prime };\omega \right) d{\bf %
r}^{\prime }  \label{rho_linear_response} \\
\Pi \left( {\bf r},{\bf r}^{\prime };\omega \right) &=&e^{2}\frac{dn}{d\mu }%
\left( -\delta ({\bf r}-{\bf r}^{\prime })+\frac{i\omega }{D_{D}}d\left( 
{\bf r},{\bf r}^{\prime };\omega \right) \right)  \label{Pi}
\end{eqnarray}
Together with the Poisson's equation, ${\bf \nabla }^{2}\phi =-4\pi \rho $,
eq. (\ref{rho_linear_response}) is an integro-differential equation with
respect to $\phi $. At small frequencies, $\omega \ll D/L^{2}$, it can be
solved by successive iterations since, in respective orders of $\omega $, it
reduces to differential equations (this point is further elaborated in
Appendix B since it is related to derivations in the quantum limit in Refs.%
\cite{LMS} - \cite{BM}).

It follows from eqs. (\ref{D_full}), (\ref{j_vs_P}) and (\ref
{j_linear_response}) that 
\begin{eqnarray}
D_{\alpha }\left( {\bf r};\omega \right) &=&\int \epsilon _{\alpha \beta
}\left( {\bf r},{\bf r}^{\prime };\omega \right) E_{\beta }\left( {\bf r}%
^{\prime };\omega \right) d{\bf r}^{\prime }  \label{D_linear_response} \\
\widehat{\epsilon }_{\alpha \beta }\left( {\bf r},{\bf r}^{\prime };\omega
\right) &=&\delta _{\alpha \beta }\delta ({\bf r}-{\bf r}^{\prime })-\frac{%
4\pi i}{\omega }\sigma _{\alpha \beta }\left( {\bf r},{\bf r}^{\prime
};\omega \right)  \label{epsilon_tensor}
\end{eqnarray}
where $\widehat{\epsilon }$ is the nonlocal dielectric tensor, which
satisfies the boundary conditions from (\ref{d_boundary_condition}). The
electric displacement vector ${\bf D}$ outside the sample coincides with the
electric field, ${\bf D=E}^{out}$, and $D_{n}$ is continuous through the
boundary. For ellipsoidal geometries (see\cite{LL}), $D_{\alpha }{\bf =}%
a_{\alpha \beta }E_{0\beta }$ (as shown above, $a_{\alpha 3}=\delta _{\alpha
3}$ and $a_{\alpha \beta }=\delta _{\alpha \beta }\left( 1+\frac{2{\cal P}}{%
E_{0}a^{3}}\right) $ for a slab and a sphere respectively). Then it is
possible to invert eq. (\ref{D_linear_response}) 
\[
E_{\alpha }\left( {\bf r};\omega \right) =const\left( \int \widehat{\epsilon 
}_{\alpha \beta }^{-1}\left( {\bf r},{\bf r}^{\prime };\omega \right) d{\bf r%
}^{\prime }\right) E_{0\beta } 
\]
so that one can find the field inside the sample as a response to external
field. A straightforward calculation for a slab (see Appendix C) confirms
that the nonlocal-$\widehat{\epsilon }$ result coincides with that obtained
in the phenomenological derivation.

The correspondence between the integral equations (\ref{j_linear_response})-(%
\ref{d_boundary_condition}) and the phenomenological description in Section
II is easily shown by applying the ${\bf \nabla }$-operator to both sides of
eq. (\ref{rho_linear_response}) and using eq. (\ref{j_linear_response_2}).
This gives the\ ''generalized Einstein transport equation'' (\ref
{Einstein_rel}) 
\[
{\bf j}\left( {\bf r};\omega \right) =\sigma _{D}{\bf E}\left( {\bf r}%
;\omega \right) -D_{D}{\bf \nabla }\rho \left( {\bf r};\omega \right) 
\]
while the boundary condition (\ref{d_boundary_condition}) is equivalent to $%
j_{n}|_{\partial }=0$. Clearly, our response function formalism correctly
reproduces the combined effect of Maxwell's equation (screening),
continuity, and diffusive approximation.

The diffuson can be written as

\begin{equation}
d\left( {\bf r},{\bf r}^{\prime };\omega \right) =\sum_{m=0}^{\infty }\frac{%
\Omega _{m}\left( {\bf r}\right) \Omega _{m}\left( {\bf r}^{\prime }\right) 
}{\lambda _{m}^{2}+\frac{i\omega }{D_{D}}}  \label{d_eigenfunctions}
\end{equation}
in terms of the volume-normalized eigenfunctions $\Omega _{m}$, such that 
\begin{eqnarray}
{\bf \nabla }^{2}\Omega _{m}\left( {\bf r}\right) +\lambda _{m}^{2}\Omega
_{m}\left( {\bf r}\right) &=&0  \label{Omega_m_equation} \\
{\bf \nabla }_{n}\Omega _{m}\left( {\bf r}\right) |_{\partial } &=&0
\label{Omega_m_bc}
\end{eqnarray}
We remark in passing on zero mode, $\lambda _{0}=0$, for which $\Omega
_{0}=V^{-1/2}$. This introduces a term 
\[
V^{-1}\int \phi \left( {\bf r};\omega \right) d{\bf r} 
\]
in the expression of $\rho $ versus $\phi $, which can be always chosen
zero. Dropping the zero mode replaces the diffusion propagator by 
\[
d\left( {\bf r},{\bf r}^{\prime };\omega \right) =\sum_{m\neq 0}^{\infty }%
\frac{\Omega _{m}\left( {\bf r}\right) \Omega _{m}\left( {\bf r}^{\prime
}\right) }{\lambda _{m}^{2}+\frac{i\omega }{D_{D}}} 
\]
The modified diffusion propagator satisfies the following equation

\begin{equation}
{\bf \nabla }^{2}d\left( {\bf r},{\bf r}^{\prime };\omega \right) =-\delta (%
{\bf r}-{\bf r}^{\prime })+\frac{1}{V}+\frac{i\omega }{D_{D}}d\left( {\bf r},%
{\bf r}^{\prime };\omega \right)  \label{d_equation_modified}
\end{equation}
and is the Fourier transform of the time-dependent diffuson of Ref.\cite{GE}
.

It must be emphasized that the response-function formalism is, obviously,
subject to the same condition of applicability - an assumption of diffusion
approximation - as the phenomenological approach.

\section{Discussion}

We have obtained closed form expressions for the absorption of the
quasi-static longitudinal electric field in a slab and sphere. We have also
derived corresponding real space expressions for the nonlocal linear
response functions. It is of interest to compare our present results with
the absorption in a transverse field. As pointed out in Ref.\cite{LL}, in a
metal the magnetic component of the transverse field is much larger than the
electric field. Therefore, we will address the magnetic dipole absorption.

It can be shown that the magnetic moment ${\cal M}$ of a slab and a sphere%
\cite{LL} in a field $H_{y}=H_{0}e^{i\omega t}$ are given respectively by 
\[
{\cal M}=-\frac{VH_{0}}{8\pi }\left( 1-\frac{\tanh \left( L\varkappa
/2\right) }{L\varkappa /2}\right) 
\]
and 
\[
{\cal M}=-\frac{VH_{0}}{8\pi }\left( 1-3\frac{\coth \left( a\varkappa
\right) }{a\varkappa }+\frac{3}{\left( a\varkappa \right) ^{2}}\right) 
\]
where 
\[
\varkappa =\frac{1+i}{\delta },\text{ }\delta =\frac{c}{\sqrt{2\pi \sigma
_{0}\omega }} 
\]
and $\delta $ is the penetration (skin) depth. Functionally, these results
are virtually identical to the ones for the longitudinal filed. The main
differences are the absence of the depolarization factors and the purely
dynamical nature of $\delta $. At low frequency the latter can be much
larger than the sample size, e.g. $a\varkappa \ll 1$, in contrast with $L%
\widetilde{\Lambda }\gg 1$ for a good metal in a longitudinal field. This
means that the transverse field can easily penetrate the metal and, as a
result, the magnetic dipole absorption 
\[
Q_{{\cal M}}=\frac{1}{2}%
\mathop{\rm Re}%
\left\{ i\omega {\cal M}H_{0}\right\} =-\frac{1}{2}\omega H_{0}%
\mathop{\rm Im}%
\left\{ {\cal M}\right\} 
\]
can become dominant\cite{LL}. Indeed, for a spherical metal particle in a
wave, $H_{0}=E_{0}$, we find that\cite{LL} 
\[
\frac{Q_{{\cal M}}}{Q_{RD}}=\frac{\pi ^{2}}{90}\left( \frac{2a\ell }{\lambda
_{p}\Lambda ^{-1}}\right) ^{2} 
\]
where $\lambda _{p}$ is the plasma wave-length\cite{AM}. Assuming $\omega
\tau \sim .05$, so that $\delta =\lambda _{p}/\sqrt{2\pi ^{2}\omega \tau }
\sim \lambda _{p}$ where $\lambda _{p}\sim 3000\stackrel{\circ }{A}$\cite{AM}
, and taking $\Lambda ^{-1}\sim 1\stackrel{\circ }{A}$\cite{AM} and $\ell
\sim 100\stackrel{\circ }{A}$ (mean-free-path is typically smaller in small
particles than in bulk materials), we find $Q_{{\cal M}}/Q_{RD}\sim 10^{2}$
for $a\sim 500\stackrel{\circ }{A}$ .

\section{Acknowledgments}

We benefitted from very useful discussions with Young Kim, Michael Ma, and
Alexander Zyuzin. This work was not supported by any funding agency.

\appendix 

\section{Absorption in terms of electric moment}

Since there is no charge outside the sample, the integral in eq. (\ref
{Q_total-1}) can be extended to the entire space. We can write 
\[
\rho \left( {\bf r};\omega \right) =-{\bf \nabla }^{2}\widehat{\phi }\left( 
{\bf r};\omega \right) 
\]
where the potential 
\[
\widehat{\phi }\left( {\bf r};\omega \right) =\phi \left( {\bf r};\omega
\right) {\bf -}\phi _{0}\left( {\bf r};\omega \right) 
\]
is such that $\widehat{\phi }\left( {\bf r\rightarrow \infty };\omega
\right) =0$. Using these equations, we find 
\begin{equation}
Q=%
\mathop{\rm Re}%
\left\{ -\frac{i\omega }{2}\int \rho \left( {\bf r};\omega \right) \phi
_{0}\left( {\bf r}\right) d{\bf r}\right\} +%
\mathop{\rm Re}%
\left\{ \frac{i\omega }{2}\int {\bf \nabla }^{2}\widehat{\phi }\left( {\bf r}%
;\omega \right) \widehat{\phi }^{*}\left( {\bf r};\omega \right) d{\bf r}%
\right\}  \label{Q_total-1-app}
\end{equation}
By the Green's formula, we find 
\begin{eqnarray*}
\mathop{\rm Re}%
\left\{ \frac{i\omega }{2}\int {\bf \nabla }^{2}\widehat{\phi }\left( {\bf r}%
;\omega \right) \widehat{\phi }^{*}\left( {\bf r};\omega \right) d{\bf r}%
\right\} &=&%
\mathop{\rm Re}%
\left\{ \frac{i\omega }{2}\int {\bf \nabla }^{2}\widehat{\phi }^{*}\left( 
{\bf r};\omega \right) \widehat{\phi }\left( {\bf r};\omega \right) d{\bf r}%
\right\} \\
&=&-%
\mathop{\rm Re}%
\left\{ \left( \frac{i\omega }{2}\int {\bf \nabla }^{2}\widehat{\phi }\left( 
{\bf r};\omega \right) \widehat{\phi }^{*}\left( {\bf r};\omega \right) d%
{\bf r}\right) ^{*}\right\} =-%
\mathop{\rm Re}%
\left\{ \left( \frac{i\omega }{2}\int {\bf \nabla }^{2}\widehat{\phi }\left( 
{\bf r};\omega \right) \widehat{\phi }^{*}\left( {\bf r};\omega \right) d%
{\bf r}\right) \right\}
\end{eqnarray*}
which shows that the second term in eq. (\ref{Q_total-1-app}) is zero and
proves eq. (\ref{Q_total-2}).

\section{Small-frequency expansion}

In the zero's order in $\omega $, we find from eq. (\ref{rho_linear_response}
) 
\begin{equation}
\rho _{st}\left( {\bf r}\right) =-e^{2}\frac{dn}{d\mu }\phi _{st}\left( {\bf %
r}\right)   \label{rho_static_linear_response}
\end{equation}
and 
\begin{equation}
\nabla ^{2}\phi _{st}\left( {\bf r}\right) -\Lambda ^{2}\phi _{st}\left( 
{\bf r}\right) =0  \label{phi_static_equation}
\end{equation}
This must be supplemented by electric field matching at the boundary.
Obviously, ${\bf j}_{st}=\sigma _{0}{\bf E}_{st}-D{\bf \nabla }\rho _{st}=0$%
. In the next order in $\omega $, $\phi =\phi _{st}+\phi _{\omega }$ and $%
\rho =\rho _{st}+\rho _{\omega },$ with $\phi _{\omega },$ $\rho _{\omega
}\propto \omega $. Using eqs. (\ref{rho_linear_response}) and (\ref
{phi_static_equation}), we find 
\begin{equation}
\rho _{\omega }\left( {\bf r};\omega \right) =-e^{2}\frac{dn}{d\mu }\left[
\phi _{\omega }\left( {\bf r}\right) -\phi _{RD}\left( {\bf r};\omega
\right) \right]   \label{rho_omega_linear_response}
\end{equation}
and the differential equation for $\phi _{\omega }$%
\begin{equation}
\nabla ^{2}\phi _{\omega }\left( {\bf r}\right) -\Lambda ^{2}\phi _{\omega
}\left( {\bf r}\right) =-\Lambda ^{2}\phi _{RD}\left( {\bf r};\omega \right) 
\label{phi_omega_equation}
\end{equation}
which is easily solvable for simple geometries, such as slab and sphere, and
which produces results corresponding to the small-frequency expansions of
the respective complete solutions.

Above, we introduced the notation 
\begin{equation}
\phi _{RD}\left( {\bf r};\omega \right) =\frac{i\omega }{D}\int d\left( {\bf %
r},{\bf r}^{\prime };0\right) \phi _{st}\left( {\bf r}^{\prime }\right) d%
{\bf r}^{\prime }  \label{phi_Rayleigh_Drude}
\end{equation}
which we call the Rayleigh-Drude potential. It gives the dynamical field in
the bulk and is responsible for absorption. As follows from eq. (\ref
{d_boundary_condition}), this field falls off to zero at the boundary, ${\bf %
E}_{RD,n}|_{\partial }=0$. In a slab, for instance, 
\begin{equation}
\phi _{RD}(z{\bf ;}\omega )=-\frac{i\omega }{4\pi \sigma _{0}}\left( \left(
z-\frac{L}{2}\right) E_{0}+\phi _{st}(z)\right) 
\label{phi_Rayleigh_Drude-slab}
\end{equation}
where we used 
\begin{equation}
d(z,z^{\prime };0)=\sum_{m=1}^{\infty }\frac{2}{L}\frac{\cos \left( \frac{m%
\pi z}{L}\right) \cos \left( \frac{m\pi z^{\prime }}{L}\right) }{\left( 
\frac{m\pi }{L}\right) ^{2}}=\frac{1}{2}\left( \frac{z^{2}}{L}-\left|
z-z^{\prime }\right| +\frac{L}{2}-z\right)   \label{d_static-slab}
\end{equation}
Comparing with expression (\ref{phi_solution-slab}), we observe that eq. (%
\ref{phi_Rayleigh_Drude-slab}) is obtained by expanding the former in $%
\omega $ with the account for the $\omega $-dependence in $\epsilon _{D}$,
while ignoring such dependence in $\widetilde{\Lambda }$ (see eq. (\ref
{Lambda_complex})).

Expanding eq. (\ref{Q_total-1}) for small frequencies, we observe that 
\[
Q=%
\mathop{\rm Re}%
\left\{ -\frac{i\omega }{2}\int \left[ \rho _{\omega }\left( {\bf r}\right)
\phi _{st}\left( {\bf r}\right) -\rho _{st}\left( {\bf r}\right) \phi
_{\omega }\left( {\bf r}\right) \right] d{\bf r}\right\} 
\]
With the use of eqs. (\ref{rho_static_linear_response}) and (\ref
{rho_omega_linear_response}), the latter reduces to 
\begin{eqnarray}
Q &=&%
\mathop{\rm Re}%
\left\{ \frac{i\omega }{2}\int \rho _{st}\left( {\bf r}\right) \phi
_{RD}\left( {\bf r};\omega \right) d{\bf r}\right\}   \nonumber \\
&=&\frac{\omega ^{2}\Lambda ^{4}}{2\left( 4\pi \right) ^{2}\sigma }\int \int
\phi _{st}\left( {\bf r}\right) d\left( {\bf r},{\bf r}^{\prime };0\right)
\phi _{st}\left( {\bf r}^{\prime }\right) d{\bf r}^{\prime }d{\bf r=}\frac{%
\omega ^{2}}{2\sigma }\int \int \rho _{st}\left( {\bf r}\right) d\left( {\bf %
r},{\bf r}^{\prime };0\right) \rho _{st}\left( {\bf r}^{\prime }\right) d%
{\bf r}^{\prime }d{\bf r}  \label{Q_LS/E/BM}
\end{eqnarray}
which should be compared with expressions used in Refs.\cite{LMS} - \cite{BM}%
\footnote{%
Notice that in these works the quantum case was considered which may
introduce additional frequency dependence.} (notice also that $\phi
_{st}\left( z\right) =\left( -E_{0}z\right) g\left( z\right) $ for a slab
and $\phi _{st}\left( {\bf r}\right) =\left( -{\bf E}_{0}\cdot {\bf r}%
\right) f\left( r\right) $ for a sphere where the functions $g$ and $f$ are
easily found from eqs. \ref{phi_static_solution-slab} and \ref
{phi_hom-sphere} respectively). A quick check for a slab confirms that,
using eqs. (\ref{d_static-slab}) and (\ref{phi_static_solution-slab}), eq. (%
\ref{Q_LS/E/BM}) yields eq. (\ref{Q_omega2_solution-slab}) for $L\Lambda \gg
1$.

\section{Dielectric tensor in a slab}

For a slab, 
\begin{equation}
d(z,z^{\prime };\omega )=\sum_{m=1}^{\infty }\frac{2}{L}\frac{\cos \left( 
\frac{m\pi z}{L}\right) \cos \left( \frac{m\pi z^{\prime }}{L}\right) }{%
\left( \frac{m\pi }{L}\right) ^{2}+i\frac{\omega }{D_{D}}}
\label{d_dynamic-slab}
\end{equation}
whereof 
\begin{equation}
\widehat{\epsilon }(z,z^{\prime };\omega )=\sum_{m=1}^{\infty }\frac{2}{L}%
\left( 1+\Lambda ^{2}\frac{1}{\left( \frac{m\pi }{L}\right) ^{2}+i\frac{%
\omega }{D_{D}}}\right) \sin \left( \frac{m\pi z}{L}\right) \sin \left( 
\frac{m\pi z^{\prime }}{L}\right)  \label{epsilon_tensor_solution-slab}
\end{equation}
and 
\begin{equation}
\widehat{\epsilon }^{-1}(z,z^{\prime };\omega )=\sum_{m=1}^{\infty }\frac{2}{%
L}\left( 1+\Lambda ^{2}\frac{1}{\left( \frac{m\pi }{L}\right) ^{2}+i\frac{%
\omega }{D_{D}}}\right) ^{-1}\sin \left( \frac{m\pi z}{L}\right) \sin \left( 
\frac{m\pi z^{\prime }}{L}\right)
\label{epsilon_tensor_inverse_solution-slab}
\end{equation}
The field inside the slab is given by

\begin{eqnarray}
E(z;\omega ) &=&E_{0}\int \widehat{\epsilon }^{-1}(z,z^{\prime };\omega
)dz^{\prime }  \label{E_versus_external-slab} \\
&=&E_{0}\frac{4}{\pi }\sum_{m-odd}^{\infty }\frac{1}{m}\left( 1+\Lambda ^{2}%
\frac{1}{\left( \frac{m\pi }{L}\right) ^{2}+i\frac{\omega }{D_{D}}}\right)
^{-1}\sin \left( \frac{m\pi z}{L}\right)  \nonumber
\end{eqnarray}
which, upon summation, yields eq. (\ref{E_solution-slab}).

\end{document}